\documentclass[onecolumn]{aastex631}

\usepackage{float}

\received{\today}
\revised{\today}
\accepted{\today}
\submitjournal{JAAVSO}
\shorttitle{UU Cet, UW Gru, and W Tuc}
\shortauthors{Parker et al.}

\begin{document}

\title{Distances for the RR Lyrae stars UU Ceti, UW Gruis, \& W Tucanae}
\correspondingauthor{Timothy Banks}
\email{tim.banks@nielsen.com}

\author{Ross Parker}
\affiliation{Department of Physical Science and Engineering, Harper College, 1200 W Algonquin Rd, Palatine, IL 60067, USA}

\author{Liam Parker}
\affiliation{Department of Physical Science and Engineering, Harper College, 1200 W Algonquin Rd, Palatine, IL 60067, USA}

\author{Hayden Parker}
\affiliation{Department of Physical Science and Engineering, Harper College, 1200 W Algonquin Rd, Palatine, IL 60067, USA}

\author{Faraz Uddin}
\affiliation{Department of Physical Science and Engineering, Harper College, 1200 W Algonquin Rd, Palatine, IL 60067, USA}

\author{Timothy Banks}
\affiliation{Department of Physical Science and Engineering, Harper College, 1200 W Algonquin Rd, Palatine, IL 60067, USA}
\affiliation{Data Science, Nielsen, 200 W Jackson, Chicago, IL 60606, USA}



\begin{abstract}

$B$, $V$, $i$, and $z$ bandpass observations were collected in late 2020 for three RRab type stars: UU Ceti, UW Gruis, and W Tucanae. The period-luminosity (PL) relationships of Catalen et al. (2004) and Caceres \& Catelan (2008) were applied to derive distances.  These were found to be in reasonable agreement with the {\em Gaia} Early DR3 distances, lending confidence to use of the PL relationships. Fourier decompositions were applied to data from the {\em TESS} space telescope to derive, using stepwise linear regression, an empirical relationship between terms of the decomposition and the pulsation period with metallicity [Fe/H].  {\em TESS} data were available for UU Cet and W Tuc out of the three studied stars. The derived equation gave metallicities in line with the literature for both stars, lending confidence to their usage in the PL-derived distances.

\end{abstract}


\keywords{Variable stars --- RR Lyrae --- Distances}


\section{Introduction} \label{sec:intro}

RR Lyrae are low-mass, horizontal branch, short period ($<1$ day), pulsating variable stars used as `standard candles' to calculate distances. They have also been used as tracers of the chemical and dynamical properties of old stellar populations within our own and nearby galaxies, and as test objects to validate theories of the evolution of low mass stars and stellar pulsation (p1, Smith, 1995). 

The European Space Agency's {\em Gaia} ({\em Gaia} Collaboration, 2018) mission provides the opportunity to compare parallax derived distances with those based on period-luminosity (PL) relationships.  Catelan~et al. (2004) showed that use of near-infrared band-passes together with PL relationships led to more reliable distance estimates than previous PL relationships, as the PL relationship becomes more linear and more tight.  Catelan et al. (2004) gave the relation for $V$ as 
\begin{equation}
   M_V = 2.288 + 0.8824 \log{Z} + 0.1079 (\log{Z})^2, 
\end{equation} 
where {\em Z} is the metallicity. Caceres~\& Catelan (2008) provided the first investigations of the RR Lyrae period-luminosity relation in the Sloan Digital Sky Survey (SDSS) system band-passes.  After a review of PL relations in various filter systems, they concluded that the $B$, $V$, $i$, and $z$ filters delivered the most promising results.   The paper confirms that redder bandpasses, specifically {\em i} and {\em z}, identify tight and simple PL relations.  The relations for {\em i} and {\em z} respectively are: 
\begin{equation}
    M_i = 0.908 - 1.035 \log{P} + 0.220 \log{Z}
\end{equation} and 
\begin{equation}
    M_z = 0.839 - 1.295 \log{P} + 0.211 \log{Z},   
\end{equation}
where {\em P} is the pulsation period in days. Equation 2 has a standard error of the estimate of 0.045 mag, and equation 3 0.037 mag. Catelan et al. (2004) do not give similar estimates for equation 1, commenting ``...for all equations presented... the statistical errors in the
derived coefficients are always very small, of order $10^{-5} - 10^{-3}$''.

The aim of this study is to firstly obtain suitable photometric observations for three RR Lyrae stars (UU Ceti, W Tucanae, and UW Gruis), apply these equations to obtain distance estimates for the stars, and compare these estimates with each other and the published distances such as from {\em Gaia}.  It is part of a wider research effort led by Dr.\ M. Fitzgerald (Edith Cowan University, Australia) investigating further RR Lyrae stars and the relation between the equations above and parallax-based distances (see, e.g., Jones, 2020; Uzpen \& Slater, 2020; Zimmerman et al, 2020; Nicolaides et al., 2021).

\begin{figure}[htbp]
    \begin{center}
    {\epsscale{0.5} \plotone{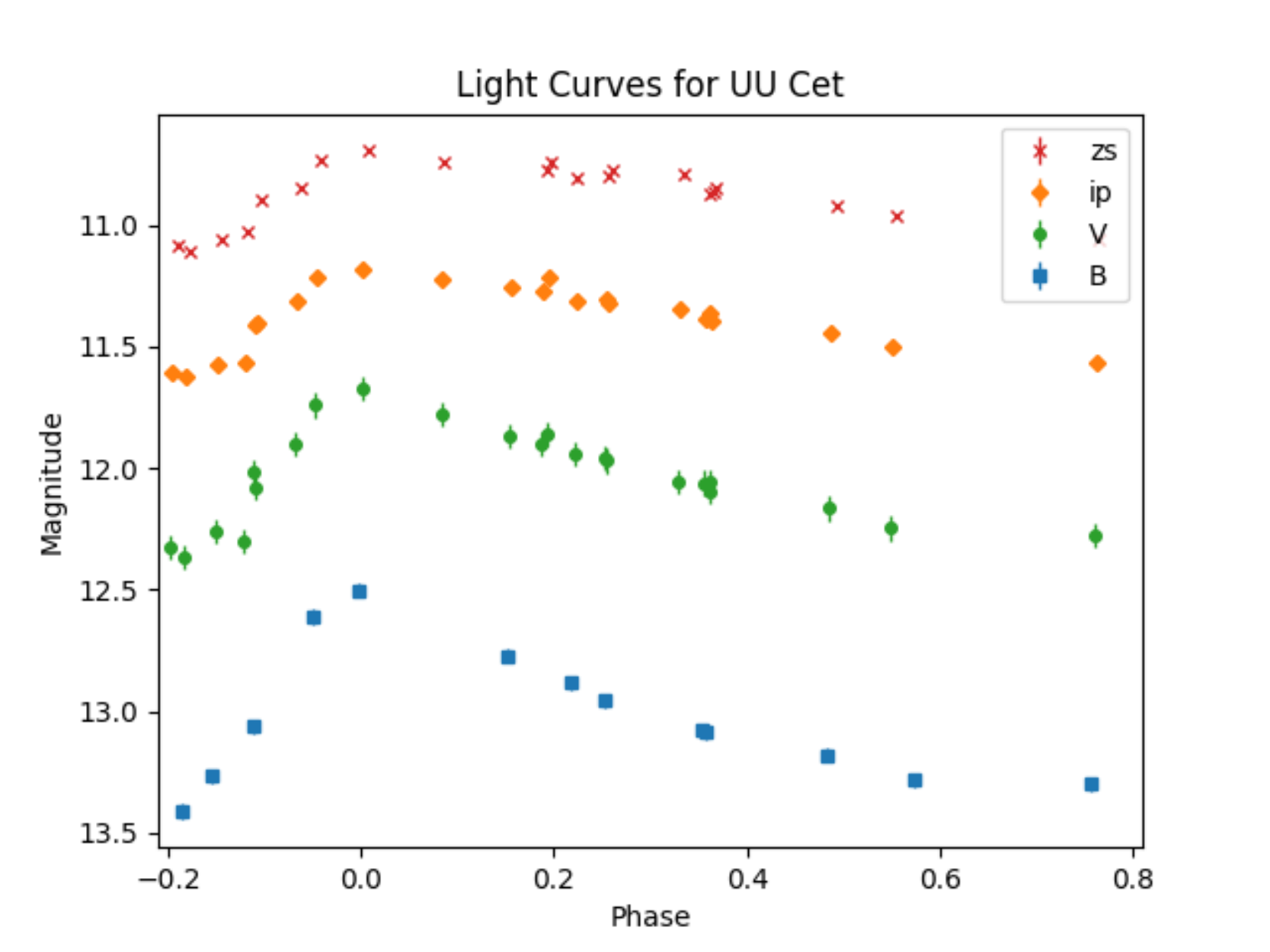}}
    \plottwo{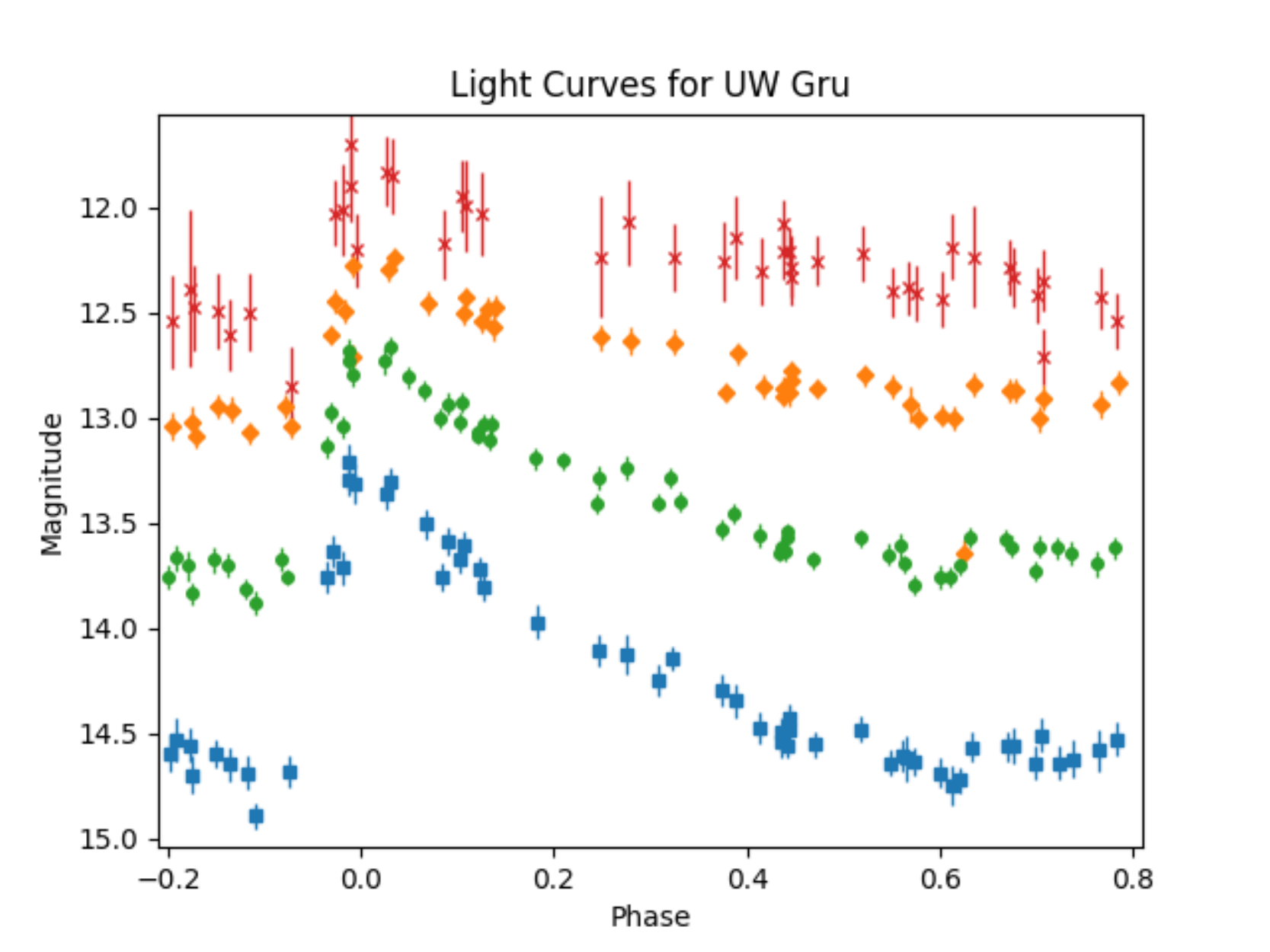}{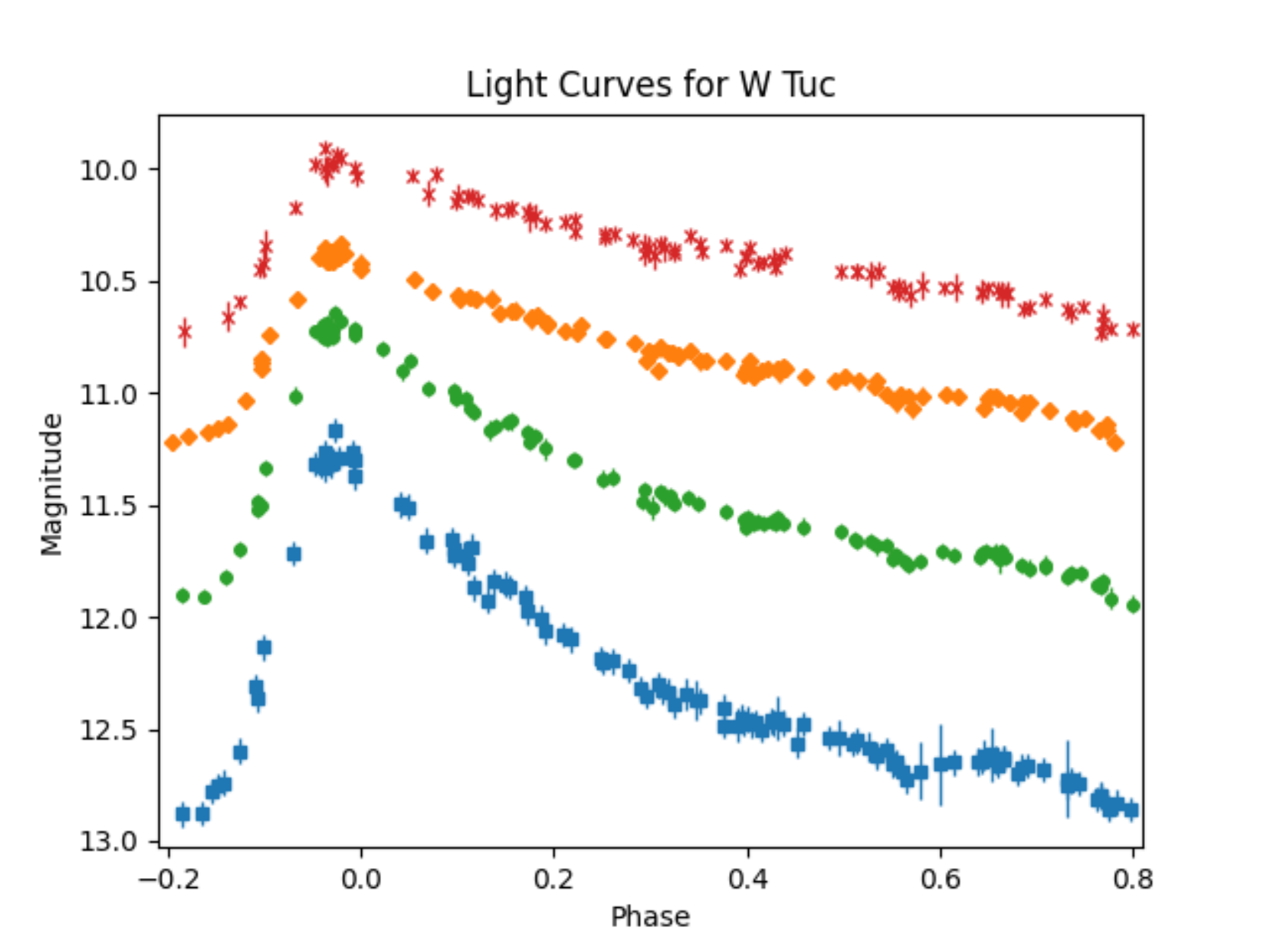}
        \caption{Light curves for the observed RR Lyrae stars. The red points correspond to $z$ band observations, orange to $i$, green to Johnson V, and blue to Johnson B. $z$, $i$, and Johnson B curves were offset by $-1$, $-0.5$ and 0.5 magnitudes respectively.}
        \label{fig:lightcurves}
        \end{center}
\end{figure}

\begin{table}[!tb]
    \caption{{\bf Exposure times} (in seconds) of the science frames for each star and filter. }
    \centering
    \hspace{-2.5cm}
    \begin{tabular}{||l|r|r|r|r||}
    \hline
    Star & $B$  & $V$ & $i$ & $z$ \\
    \hline
    UU Cet & 185 & 60 & 80 & 360 \\
    UW Gru &  60 & 30 & 60 & 240 \\
    W Tuc  &  60 & 30 & 90 & 180 \\
    \hline 
    \end{tabular} 
    \label{tab:exposure times}
\end{table}


\begin{table}[!tb]
    \caption{{\bf Calibration information:} for each star and filter, number of on-frame calibration stars is given in the column labelled `Calibration', `Frames' gives the number of processed images, `Catalog' the source of the calibration information (the reference catalog).  `SE' is the standard error (in magitudes) of the calibration as calculated by {\tt astrosource}. }
    \label{tab:data_processing}
    \centering
        \hspace{-2.5cm}
    \begin{tabular}{||l|c|c|r|c|c|c||}
    \hline
    Star   &   Filter  &  Calibration  & Frames & Catalog & SE \\
    \hline
    UU Cet & $B$ & 6 & 13  & APASS     & 0.0164 \\
           & $V$ & 7 & 23  & APASS     & 0.0093 \\
           & $i$ & 6 & 23  & Skymapper & 0.0115 \\
           & $z$ & 4 & 21  & Skymapper & 0.0247 \\
    \hline
    UW Gru & $B$ & 6 & 55  & APASS     & 0.0157 \\
           & $V$ & 6 & 63  & APASS     & 0.0105 \\
           & $i$ & 8 & 48  & Skymapper & 0.0109 \\
           & $z$ & 4 & 44  & Skymapper & 0.0083 \\
    \hline
    W Tuc  & $B$ & 7 & 108 & APASS     & 0.0139 \\
           & $V$ & 6 & 102 & APASS     & 0.0115 \\
           & $i$ & 8 & 102 & Skymapper & 0.0119 \\
           & $z$ & 7 & 93  & Skymapper & 0.0095 \\
    \hline
    \end{tabular} 
\end{table}   

\begin{table}[!tb]
    \caption{{\bf Calculated photometric data for the studied stars} (in magnitudes).
    `Max' is the maximum magnitude numerically (so the brightest for the star), `Min' the minimum, and `Mid' is the arithmetic mean of these two
    extremes.}
    \label{tab:second_table}
        \hspace{-4.5cm}
    \centering
    \begin{tabular}{||l|r|r|r |r|r|r |r|r|r||}
    \hline 
    & \multicolumn{3}{|c|}{UU Cet}  
    & \multicolumn{3}{|c|}{UW Gru} 
    & \multicolumn{3}{|c||}{W Tuc} \\
    \hline
    & Min & Max & Mid & Min & Max & Mid & Min & Max & Mid \\
    \hline
$B$ & 12.909  &  12.003 &  12.456 & 14.391  &  12.708 &  13.549 & 12.378  &  10.670 &  11.524 \\
$V$ & 12.366  &  11.672 &  12.019 & 13.679  &  12.665 &  13.272 & 11.944  &  10.641 &  11.293 \\
$i$ &  12.126 &  11.682 &  11.903 & 14.139  &  12.738 &  13.438 & 11.719  &  10.833 &  11.276 \\
$z$ & 12.108  &  11.692 &  11.900 & 13.848  &  12.704 &  13.276 & 11.729  &  10.904 &  11.317 \\
    \hline
    \end{tabular}
\end{table}


\subsection{UU Cet}

UU Cet (RRab, $\rm V_{max} = 11.688, V_{min} = 12.237$: {\em Gaia} Collaboration, 2018; $\rm V_{max} = 11.718, V_{min} = 12.350$: Clementini et al., 1992) has been documented in many different catalogs.  However only a few papers, by a research group led by Cacciari, narrow down their research to study UU Cet extensively. In Cacciari et al.\ (1992) the authors performed the Baade-Wesselink (BW) method (Baade, 1926; Wesselink, 1946) on UU Cet using previous observations from published papers.  The Infrared Flux (IF) method indicated a distance of 1887 pc, and the Surface Brightness (SB) method a distance of $1825 - 1982$ pc with values calculated with both optical colors and $(V-K)$ colors. These numbers are similar to calculations made in the paper Clementini et al.\ (1992) which also performed the BW method on UU Cet using different input variables. Parallax estimates for the star vary across researchers, as shown in Table~\ref{tab:first_table}. 

Cacciari et al.\ (1992) found metallicity to be $-1.0 \pm 0.2$, however a concrete value for [Fe/H] does not appear to have been settled on for UU Cet as it varies in the literature. For example, Chiba~\& Yoshii (1998) give an [Fe/H] value of $-1.32 \pm 0.20$ while Sandage (1993) calculated it to be $-0.79$. Cacciari et al.\ (1992) derived $0.606075^{\rm d}$ for period of UU Cet, which is very similar to other findings such as $0.60608^{\rm d }$ found by Lub (1977a) and $0.60606^{\rm d}$ from ASAS (Pojmanski, 1997).  A previous observation by Jones (1973) estimated a $(k-b)_2$ value of $0.08 \pm .019$ which is possible evidence of a Blazhko effect, however no effect was observed in the current paper (although our data are sparse, see Figure~\ref{fig:lightcurves}).


\subsection{UW Gru}

UW Gru (RRab) hasn’t often been a focal point in many papers as an object of interest. It was first discovered by Hoffmeister (1963), who classified the star as an RR Lyrae with extreme magnitudes between 12 and 13. The next publication on UW Gru was when Alain Bernard collected photoelectric $UBV$ observations over the course of three years (Bernard, 1982). The star varied between 12.6 and 13.6 in $V$ (see Figure 1 of Bernard, 1982). The Wide-field Infrared Survey Explorer all-sky mission (Gavrilchenko et al., 2014) lists UW Gru’s period at $0.5650 \pm 0.0070$ days and a distance of $3282 \pm 64$ parsecs. The distance and uncertainty were calculated using a mid-IR period–luminosity relation.  The WISE period was similar to the period of $0.548210^{\rm d}$ found in Bernard \& Burnet (1982).  Additionally, the [Fe/H] was estimated at $ -1.6 \pm  0.2$ (Bernard, 1982) and listed at $-1.41$ dex metallicity on a common [Fe/H] scale (Jercsik~\& Kovacs, 1996), which reflects the findings of other authors. Blazhko behavior wasn’t considered a factor for UW Gru, which we confirm (see Figure~\ref{fig:lightcurves}). The distance was measured at R = $2900 \pm 250$ pc from the sun, and 2550 pc below the galactic plane (Bernard~\& Burnet, 1982), although this was based on an assumed absolute magnitude. However, the given parallax from the {\em Gaia} Data Releases all differ (see Table~\ref{tab:first_table}).

\subsection{W Tuc}

While W Tuc (RRab, $\rm V_{max} = 10.96, V_{min} = 12.03$: Torrealba et al., 2015) is present in many catalogued results for RR Lyrae stars, only a handful of papers have focused on this star as a specific object of interest.  These were written primarily by a group led by Cacciari.  Cacciari et al.\ (1992)  presented JHK light curves for W Tuc, using these together with literature data, such as CORAVEL radial velocities and BVRI photometry (from Cacciari et al., 1987; Clementini et al., 1990), and the BW method to derive absolute parameters for the star. Using surface brightness methods gave a distance of 1601 to 1667 parsecs (using optical and ($V - K$) colors, $\sim 0.625$ mas) while infrared fluxes indicated a distance of 1555 pc ($\sim 0.643$ mas).  [Fe/H] was estimated as $-1.50$ ($\sigma = 0.25$).  No Blazhko effect was evident, which we confirm in this paper (see Figure~\ref{fig:lightcurves}).  Cacciari et al.\ (1992) calculated an ephermides of $2447490.719 + (0.642235 \times N)$ days, where $N$ is the cycle number.  The period is not substantially different from the 0.6422299 days as given by both Kukarin et al. (1970) and Lub (1977b). The Wide-field Infrared Survey Explorer all-sky mission (Gavrilchenko et al., 2014) lists W Tuc's period as $0.5990 \pm 0.0040$ days and a distance of $1514 \pm 20$ parsecs ($\sim 0.660$ mas).  Distance estimates for W Tuc in the literature are quite variable, as shown in Table~\ref{tab:first_table} and the distances mentioned above, although there appears to be more of an agreement towards a parallax of $\sim 0.6$ rather than $\sim 3-5$ mas. Feast et al.\ (2008) provided a later [Fe/H] estimate of $-1.57$ solar, along with $-1.76$ from Marsakov et al.\ (2018) and $-1.76$ from Dambis et al.\ (2013).  


\section{Method}

$B$, $V$, $i$, and $z$ observations were collected for these three systems using the Las Cumbres Observatory (LCO, Brown et al., 2013) automated 0.4-m SBIG telescopes over a five month period (August 2020 to December 2020).  This is the first time these stars have been observed using the $i$ and $z$ filters.  Up to three or four sets of observations were taken each night, depending on the automated scheduled and observing loads of the network. Exposure times are given in Table~\ref{tab:exposure times}. Five different observing sites inside the LCO network were used, namely Siding Springs (Australia), Sutherland (South Africa), Cerro Tololo (Chile), Haleakala (Maui), and Teide (Spain). The resulting images were processed through the OurSolarSiblings (OSS) data reduction pipeline (Fitzgerald, 2018).  OSS performs basic processing such as flat-fielding and cosmic ray removal.  These data were then input to the {\tt astrosource} software which processed the photometry of the target and comparison stars (see Fitzgerald, 2018, and Fitzgerald et al., 2021, for further details on this software). {\tt astrosource}\footnote{Version 1.5.2 is available from https://pypi.org/project/astrosource/} has the following procedure:
\begin{itemize}
\item{It first identifies stars having sufficient signal-to-noise (within the linear range of the imager) which are in all the frames being processed for a given filter.}
\item{Next the variability of these identified stars is calculated in order to identify a subset of the least variable stars, which will be used as the final ensemble set of comparison stars. Sarva et al.\ (2020) explain the selection of comparison stars by {\tt astrosource}.  First, the flux of all the potential comparison stars is summed up as if to create a single comparison `star'.  Then the variability of each comparison star across the observations is compared  with the variability of this sum across the same observations. A candidate star with variability greater than 3 times the standard deviation of the combined variability is removed from consideration.  This process loops until the variability of the combined `star' is less than or equal to 0.002 magnitudes. The remaining stars are then used as comparison stars for the data reduction,  leading to differential photometry of the target star against them. It is possible than the process ends here if no suitable comparison stars are found.  The standard errors from this process are reported in Table~\ref{tab:data_processing}, for each of the stars analyzed in this paper.}
\item{The ensemble set of known stars in the field is calibrated using APASS (Henden, Levine,
Terrell, \& Welch, 2015), SDSS (Alam et al., 2015), PanSTARRS (Magnier et al., 2016) or Skymapper (Wolf et al., 2018) depending on filter selection and declination. Fitzgerald (2018) gives further details on the calibration equations (which include color correction, extinction, and the possibility of time dependent terms if needed), making
use of the generalized method for observations across multiple nights as outlined by Harris et al.\ (1981).}
\item{The software extracts and outputs the photometric estimates, together with diagnostics and charts. Methods based on aperture and point-spread functions (e.g., DAOPhot; Stetson, 1987) are available.  After testing several methods, for this project the SEK (Source Extractor: Kron; Bertin, \& Arnouts, 1996) method was found to produce calibration estimates with the least variance.}
\item{Finally, {\tt astrosource} calculates periods using the Phase-Dispersion Minimization (PDM) and String-Length algorithms (Altunin, Caputo, \& Tock, 2020)}
\end{itemize}
Information on the number of calibration stars is provided in Table~\ref{tab:data_processing}, along with the number of science frames, the reference catalogs used, and measures of photometric accuracy. Table~\ref{tab:second_table} gives the calibrated magnitudes for the three targets stars, along with the errors as estimated by {\tt astrosource}. Our photometric data for all three stars has been uploaded to and is available in the AAVSO catalog.



\begin{table}[!tbp]
    \caption{{\bf Literature parallaxes (milli-arcseconds) for the studied stars:} ESA (1997) was the
    original data release for the HIPPARCOS mission, followed by van Leeuwen's (2007)
    revisions.  Clearly the selected stars are outside the reliable range of HIPPARCOS.
    The different {\em Gaia} Data Releases (DR) are described in 
    {\em Gaia} Collaboration (2016, 2018, 2021). EDR is an early data release, ahead of the
    later formal one.  DR3 is pending.}
    \label{tab:first_table}
    \centering
        \hspace{-1.5cm}
    \begin{tabular}{||l|c|c|c||}
    \hline
         Parallax (mas) &  UU Cet              & UW Gru                      & W Tuc \\
    \hline
         ESA            & $6.48 \pm 4.13$      &                             &  $4.88 \pm 1.88$ \\
         van Leeuwen    & $5.30 \pm 4.06$      &                             &  $3.33 \pm 1.53$ \\
         Gaia DR1       &                      &  $0.0678 \pm 0.2287$        &  $0.720 \pm 0.250$ \\
         Gaia DR2       &  $0.3823 \pm 0.0044$ & $0.2886 \pm 0.0204$         &  $0.5657 \pm 0.0256$ \\
         Gaia EDR3       &   $0.493 \pm 0.0191$ & $0.3299 \pm 0.0158$         &  $0.5963 \pm 0.0133$ \\
    \hline
    \end{tabular}
\end{table}




\begin{figure}[htbp]
    \plottwo{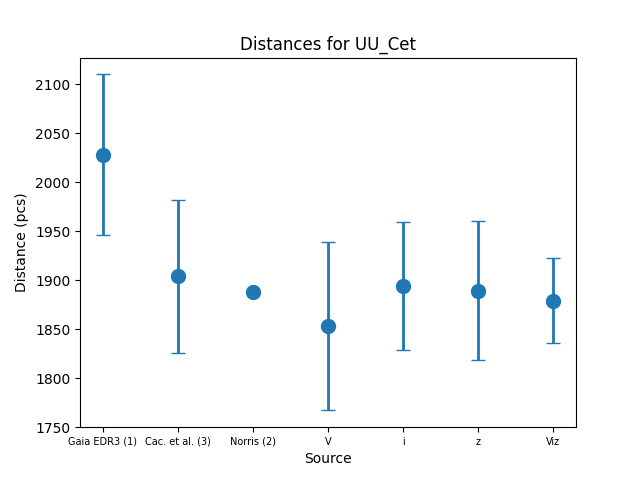}{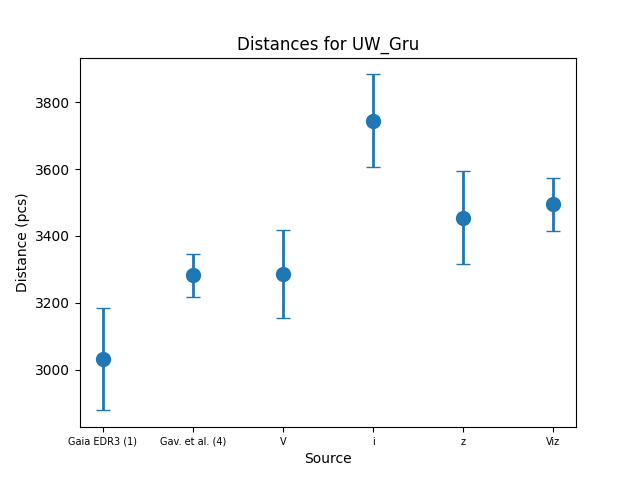}
    {\epsscale{0.5}  \plotone{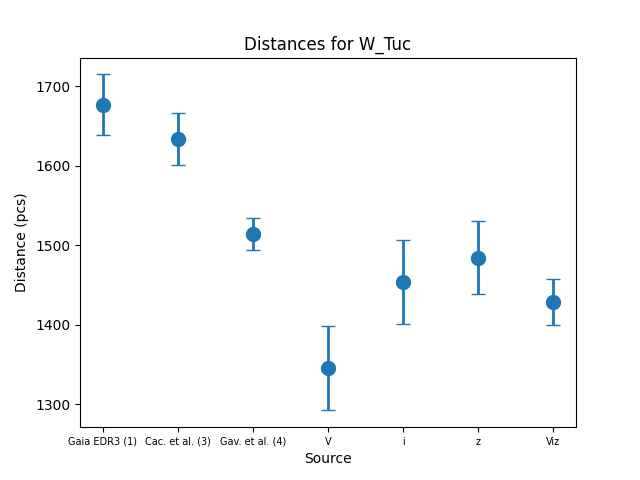}}
    \caption{{\bf Distance comparisons for UU Cet, UW Gru, and W Tuc} Some sources are indicated by number for space reasons: (1) {\em Gaia} Collaboration, 2021, (2) Norris, 1986, who used Hemenway's (1975) statistical parallax calculation, (3) distances based on the Surface Brightness method by Cacciari et al., 1992, and (4) Gavrilchenko et al., 2014, who used a mid-IR period–luminosity relation. No error value was given for Norris (1986). EDR3 refers to the early data release 3.}
        \label{fig:distances}
\end{figure}

\begin{figure}[htbp]
     \centering
     \includegraphics[width=12cm, height= 8cm]{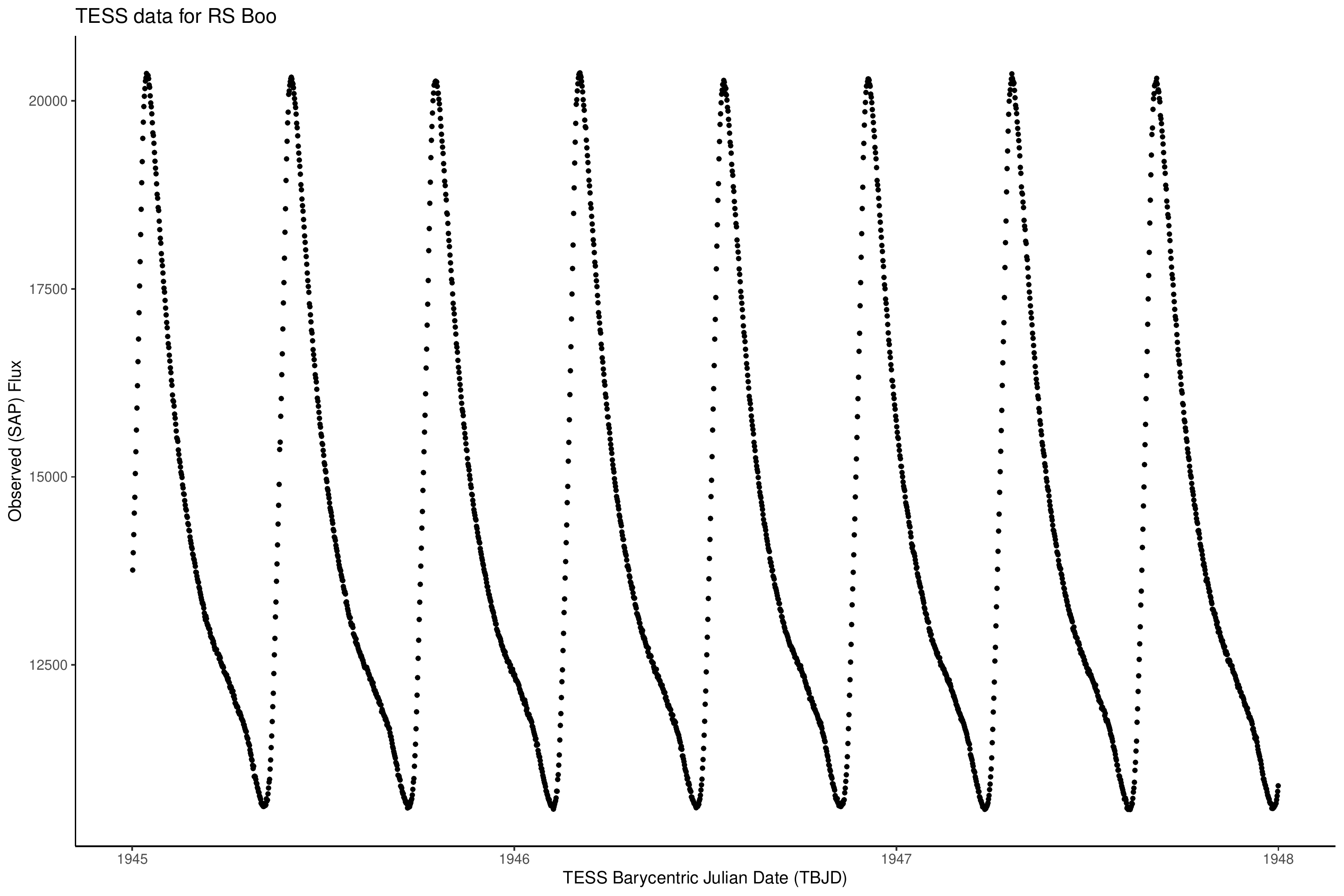}
    \caption{{\bf {\em TESS} observations of RS Boo:} shows the high precision that the mission is 
    capable of.  Fluxes were derived using the {\em TESS} standard `simple aperture photometry' (SAP) data pipeline, and are plotted against {\em TESS} Barycentric Julian Date (add 2457000 for `normal'
    Julian Dates). }
     \label{fig:rs_boo}
\end{figure}

\begin{figure}[htbp]
     \centering
     \includegraphics[width=12cm]{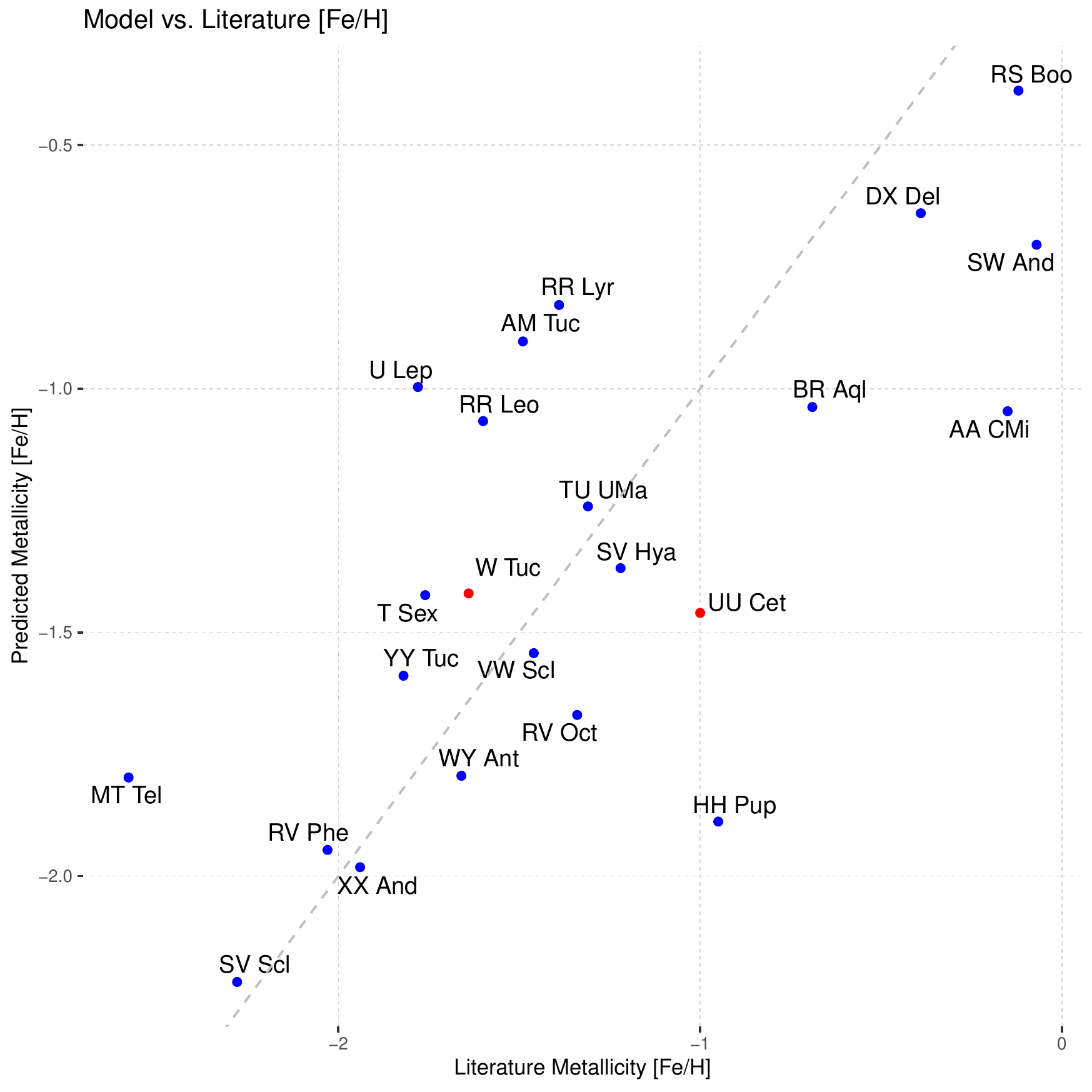}
     \caption{{\bf Model vs. Literature Metallicities} for an arbitrary selection of RR Lyrae stars
     observed by the {\em TESS} mission. The literature metallicity is on the horizontal axis, while the
     modelled metallicity is on the vertical.  The dotted grey line is that of perfect agreement 
     between the literature and the model. The model predictions and literature values for UU Cet and 
     W Tuc are indicated by the red dots.  All other systems were used to train the model.}
     \label{fig:mast_metals}
\end{figure}



\begin{table}[!tbp]
    \caption{{\bf Input parameters for the studied stars:} Period is average calculation from the 4 light bands in the current study, using PDM-based estimates. The extinction factor for UU Cet is an average of 3 values taken from previous observations: [1] Cacciari et al. (1992), [2] Feast et al. (2008), and [3] Schlafly \& Finkbeiner (2011). [4] indicates Cacciari et al. (1992) as the source of the information, [5] Jurcsik \& Kovacs (1996). $ \log{Z}$ were calculated from the [Fe/H] values supplied. The metallicities from the compilation of Dambis et al. (2013) are not dramatically different for the three stars, being $-1.32$, $-1.68$, and $-1.64$ for UU Cet, UW Gru, and W Tuc respectively, nor from the metallicities in Table 7 for UU Cet and W Tuc. }
    \label{tab:third_table}
    \centering
        \hspace{-1.5cm}
    \begin{tabular}{||l|c|c|c||}
    \hline
                    & UU Cet               & UW Gru                             & W Tuc \\
    \hline
         Period     & $0.60608 \pm 0.0055$      &   $0.548375 \pm 0.000625$     &  $0.642233 \pm 0.00086$  \\
         $E(B-V)$   & $0.022$   [1,2,3]         &   $0.021$ [3]                 &  $0.02$ [3] \\
         $[Fe/H]$   &  $-1.0 \pm 0.2$ [4]       &    $-1.41$ [5]                &  $-1.57$ [2] \\
         $\log{Z}$  &         $-2.551$          & $-2.961$                      &  $-3.121$  \\
    \hline
    \end{tabular}

\end{table}

\begin{table}[!tbp]
\centering
\caption{{\bf Calculated distances} for the target stars, in parsecs. The mean is across
    the three filters $V$, $i$, and $z$.}
\label{tab:distances}
    \hspace{-1.5cm}
\begin{tabular}{||l|c|c|c|c||}
\hline
Star   & $V$ & $i$ & $z$ & Mean ($Viz$)\\
\hline
UU Cet &  $1746 \pm 89$  & $1808 \pm 63$  & $1807 \pm 69$  &  $1787 \pm 42$ \\
UW Gru &  $3287 \pm 132$ & $3745 \pm 138$ & $3454 \pm 139$ &  $3495 \pm 79$ \\
W Tuc  &  $1345 \pm 53$  & $1454 \pm 53$  & $1484 \pm 46$  &  $1428 \pm 29$ \\
\hline
\end{tabular}
\end{table}


\section{Results}

Our data for UW Gru agree well with the $BV$ photometry of Bernard (1982), covering the same ranges bar that our $B$ data covered the dip (just before the star brightens again) which was not covered by Bernard and so we have a fainter magnitude limit for that band.  Similarly, our UU Cet data are in good agreement with Clementini et al. (1992). This paper's $V$ range of [10.64, 11.94] for W Tuc is lower than Torealba et al.'s (2000) range of [10.96, 12.03] as well as the intensity mean of 11.43 from the literature compilation of Dambis et al. (2013, compared to our mean magnitude of 11.29). However we note Figure 1b of Clementini et al. (1990), which plots light curves (and colors) of W Tuc, and Table IIIb of the same paper which show a $V$ light curve varying between 10.78 and 11.95, are in closer agreement with our estimates. Figure 1(c) shows that our lowest magnitude (brightest) is set by a single point, with nearby phases being more dim, bringing our photometry closer to Clementini et al. (1990). Binning the data would have reduced the impact of such apparent outliers.  This was not attempted as there were few observations around the peak phase.  Perhaps taking the peak magnitude from the Fourier analyses (below) would have been more robust, provided there are sufficient data across the cycle.

In order to calculate distances to the target stars the processed data, together with information from the literature (see Table~\ref{tab:third_table}), were used to populate the PL relations from Catelan (2004) and Caceres~\& Catelan (2008). The period estimates from this study are given in Table~\ref{tab:third_table}, and are an arithmetic mean across the four band-passes and two methods mentioned above. These values are in good agreement with the literature (see the reviews above), bar the WISE estimates, indicating no significant period changes.  The calculated distances are given in Table~\ref{tab:distances} and charted as Fig~\ref{fig:distances} (on page~\pageref{fig:distances}). In general the agreement between the preliminary {\em Gaia} EDR3 release and the estimates from this study's light curves is reasonable.  As expected, the $V$ band shows the greatest difference to {\em Gaia}, indicating closer distances.  Distances for the other two band passes are in good agreement, bar the $i$ distance for UW Gru although it's error bar does just overlap that of the $z$ band. Pop~\& Richney (2021) obtained observations for SX For using the LCO network and processed the data in an identical manner to this study, as did Lester et al.\ (2021) for YZ Cap.  They too found good agreement between the {\em Gaia} data and their calculated distances using the PL relations.  While it is an extrapolation, these comparisons suggest that the PL relations could be used with some confidence for distances beyond {\em Gaia's} capabilities.

Fourier decompositions of light curves have been used to derive relationships between [Fe/H], period, and some of the component sine waves (Simon, 1988; Kovacs \& Zsoldos, 1995).   The recent {\em TESS} mission (Ricker et al., 2015) has provided very high accuracy photometry of a number of RR Lyrae stars during its survey (see Figure~\ref{fig:rs_boo} on page~\pageref{fig:rs_boo} for an example), which we used to build such a relationship and apply it to W Tuc and UU Cet in a check that the literature reddenings were reasonable. Both Simon (1988) and Kovacs \& Zsoldos (1995) used Johnson V for their equations.  We did not feel comfortable applying these relationships to the {\em TESS} data given its band-pass covers approximately 600 to 1000 nm, and is essentially centered on the Cousins I-band (which has a central wavelength of approximately 787 nm) to the red of Johnson V (central wavelength of approx.\ 575 nm, with a full-width half maximum of approximately 99 nm). The {\em TESS} mission was designed as a planet hunter, being optimized to search M dwarfs as possible host stars.  The band-pass was chosen to reduce photon counting noise, and to increase the mission's ability to detect small planets transiting late type stars. The long wavelength band-pass end is set by the CCD detectors themselves, being their red limits, and the short wavelength end is set by a coating on the camera lenses. We therefore attempted to build a relationship for {\em TESS} observations, noting that while the {\em TESS} data were of high quality our model would be highly dependent on the quality of the [Fe/H] values used to build it. 

No 2-minute cadence prepared {\em TESS} light curves were available for UW Gru, so we were unable to fit this star.  Light curves  for the other two stars were downloaded from the Mikulski Archive for Space Telescopes (MAST, Jenkins et al, 2016). We used the straight Simple Aperture Photometry (SAP) data, applying the {\sc period04} (Lenz \& Breger, 2005) program for the 12 component decomposition which fitted the following standard equation:
\noindent
\begin{minipage}{\linewidth}
\begin{equation}
f(t)=Z+\sum_{i=1,n} A_{i} \sin \left(2 \pi\left(\Omega_{i} t+\phi_{i}\right)\right)
\end{equation}
\end{minipage}
where $t$ is time. Stepwise linear regression was conducted in R (R Core Team, 2017), using period data from the {\sc nitro9} online archive for RR Lyrae Fourier decomposition\footnote{https://nitro9.earth.uni.edu/fourier/index.html} and metallicity from the SIMBAD (Wenger et al., 2000) system summaries. The set of initial variables were the period, $\phi_{1}$ to $\phi_6$ inclusive (where $\phi_i$ is in the range between 0 and 1 phase inclusive), and $\phi_{2,1}$ to $\phi_{6,1}$ inclusive following the formula $\phi_{j, 1}=\phi_{j}-j \phi_{1}$ (in the same range). Basic data are given in Table~\ref{tab:fourier}.  W Tuc and UU Cet were not included in the training data set, which was made up of 21 stars. Regression in both directions (forwards and backwards) settled on the following equation:
\begin{equation}
    [{Fe}/H] = -2.4083 P - 1.2950 \phi_{1} + 1.2888 \phi_{5} - 1.4273 \phi_{6} + 1.341 \phi_{3,1}
\end{equation}
where $P$ is the light curve period (in days), with all terms significant at the 1\% level or better. The adjusted $R^2$ value was 0.87, indicating a good model although by eye it does seem to be over-estimating the metallicity for low [Fe/H] stars.  This model was a better fit than one including a constant term. As can be seen by the scatter about the line of perfect agreement in Figure~\ref{fig:mast_metals} (on page~\pageref{fig:mast_metals}), the standard deviation of the residuals is relatively large at 0.49.    As a comparison, the relationship derived by Kovacs~\& Zsoldos (1995) had a prediction accuracy of 0.23 to 0.18 dex.  We caution that this should be considered a pilot study, and that these promising results could be built on by a more rigorous follow-up study (although it could be that the wide band-pass itself leads to imprecision).

We then applied this model to the {\sc period04} parameters for UU Cet and W Tuc, finding reasonable agreement with the literature values (see Figure~\ref{fig:mast_metals}), increasing our confidence in the literature values used in the distance estimates for these stars, which in turn are used in the PL relationships given the relationships $[M / H]= [F e / H] \mid+\log (0.698 f+0.362)$
and $\log Z=[M / H]-1.765$, where $f = 10^{0.3}$. The derived [Fe/H] for UU Cet was $-1.46$ and that for W Tuc $-1.64$.  Both stars are in the mid-range of the data, and so less affected by questions about the model fit at the extremes. We note that the UU Cet [Fe/H] is not well constrained in the literature (see above), ranging from Chiba et al'.s (1998) value of $-1.32 \pm 0.20$ down to Sandage's (1993) value of $-0.79$. Our calculated value is at the upper end of this range. Using our model's predicted value of $-1.46$ we calculate distances that are $\sim 100$ pcs more then our current results, moving closer to the {\em Gaia} estimates. Additional stars could be included into the model building perhaps leading to an improved empirical relationship, for instance for the low metallicity stars which do not seem to be so well modelled by the current equation.  We also note that our caution about applying the equation of Kovacs~\& Zsoldos might have been misplaced, using their PL relationship gave metallicities of $-1.29$ and $-1.50$ for UU Cet and W Tuc respectively.

\newpage
\section{Summary}
Using {\tt astrosource} to perform photometric analysis, distances were derived for UU Cet, UW Gru, and W Tuc. The average between the $V$, $i$, and $z$ ($Viz$) filters were focused on to compare with the distances found in literature and {\em Gaia}. The calculated distance for UU Cet was $1787 \pm 42$ pc, UW Gru's calculated distance was $3495 \pm 79$ pc, and W Tuc's calculated distance was $1428 \pm 29$ pc. The distances of all three stars varied compared to the distances in literature and {\em Gaia}. Though the calculated distances were not in full agreement with {\em Gaia}, they were not far off and this is encouraging that further work could demonstrate a closer agreement between the applied PL equations and the parallax-based distance estimates. If shown, then this would lend support for using the PL estimates at distances beyond {\em Gaia's} `distance of reliability', noting of course that this would be an extrapolation. There was no apparent pattern found with all three stars: while UU Cet and W Tuc's distances were closer than {\em Gaia}, UW Gru's distance was further.  

There is considerable variation in literature estimates for metallicity of these stars. The choice of a given metallicity will impact the distance calculated using the PL relations. We therefore have applied Fourier analysis to the high quality light curves from the {\em TESS} space telescope, and attempted a calibration given the wide band pass used by this mission.  The calibration/model was used to derive metallicities for UU Cet and W Tuc. This led to an upwards revision of some 100 pc for the distances for these stars, bringing them closer to the {\em Gaia} distances.  We believe further work to improve the calibration is worthwhile, particularly given the near full sky coverage of {\em TESS} means a greater number of RR Lyrae stars will have been observed than have been tested in this paper.  


\begin{acknowledgments}

This work has made use of data from the European Space Agency (ESA) mission {\it Gaia} (\url{https://www.cosmos.esa.int/gaia}), processed by the {\it Gaia} Data Processing and Analysis Consortium (DPAC, \url{https://www.cosmos.esa.int/web/gaia/dpac/consortium}). Funding for the DPAC has been provided by national institutions, in particular the institutions participating in the {\it Gaia} Multilateral Agreement. We are grateful for observing time on the Las Cumbres Observatory Global Telescope Network, and to Dr.\ M. Fitzgerald for making available this opportunity. This research has made use of the International Variable Star Index (VSX) database, operated at AAVSO, Cambridge, Massachusetts, USA. It also includes data collected by the {\em TESS} mission and obtained from the MAST data archive at the Space Telescope Science Institute (STScI). Funding for the {\em TESS} mission is provided by the NASA's Science Mission Directorate.  STScI is operated by the Association of Universities for Research in Astronomy, Inc., under NASA contract NAS 5–26555. We made use of the SIMBAD database, operated at CDS, Strasbourg, France. We thank the anonymous referee for their helpful comments which improved this paper.

\end{acknowledgments}


\begin{table}[b]
    \centering
        \hspace{-2.5cm}
    \begin{tabular}{||l|r|r|r|r|r|r|r|r||}
    \hline
    Star & [Fe/H] & Period (days) & $\phi_1$ & $\phi_2$ & $\phi_3$ & $\phi_4$ & $\phi_5$ & $\phi_6$ \\
    \hline
DX Del	& -0.39	& 0.472611	& 0.35056	& 0.40746	& 0.82712	& 0.82430	& 0.30808	& 0.33452 \\
HH Pup	& -0.95	& 0.390746	& 0.81737	& 0.45289	& 0.31223	& 0.96700	& 0.06399	& 0.78401 \\
AA CMi	& -0.15	& 0.476373	& 0.99612	& 0.64010	& 0.78695	& 0.36239	& 0.89092	& 0.57894 \\
BR Aql	& -0.69	& 0.481878	& 0.38438	& 0.84380	& 0.28869	& 0.59572	& 0.40716	& 0.06225 \\
RR Lyr	& -1.39	& 0.566798	& 0.29121	& 0.28716	& 0.75002	& 0.34386	& 0.43074	& 0.56504 \\
WY Ant	& -1.66	& 0.574337	& 0.66456	& 0.31873	& 0.96257	& 0.70657	& 0.31050	& 0.86966 \\
RS Boo	& -0.12	& 0.377334	& 0.06832	& 0.41788	& 0.39094	& 0.89740	& 0.92422	& 0.58002 \\
VW Scl	& -1.46	& 0.510915	& 0.27568	& 0.23861	& 0.14502	& 0.23508	& 0.26144	& 0.50160 \\
TU UMa	& -1.31	& 0.557650	& 0.37273	& 0.58698	& 0.96831	& 0.22330	& 0.50984	& 0.84274 \\
YY Tuc	& -1.82	& 0.635020	& 0.92508	& 0.74387	& 0.53565	& 0.35213	& 0.13144	& 0.03621 \\
AM Tuc	& -1.49	& 0.405791	& 0.66800	& 0.79672	& 0.11896	& 0.62165	& 0.92133	& 0.28531 \\
MT Tel	& -2.58	& 0.316901	& 0.69575	& 0.33340	& 0.61760	& 0.81035	& 0.41737	& 0.96622 \\
T Sex	& -1.76	& 0.324680	& 0.99883	& 0.96056	& 0.40241	& 0.67227	& 0.20861	& 0.11673 \\
SV Scl	& -2.28	& 0.377359	& 0.83622	& 0.84052	& 0.84700	& 0.36359	& 0.17991	& 0.63991 \\
RV Phe	& -2.03	& 0.596419	& 0.16356	& 0.52287	& 0.68158	& 0.20848	& 0.18824	& 0.55687 \\
RV Oct	& -1.34	& 0.571158	& 0.01088	& 0.42023	& 0.23609	& 0.33863	& 0.07027	& 0.44831 \\
U Lep	& -1.78	& 0.581458	& 0.27802	& 0.38164	& 0.61095	& 0.80470	& 0.02459	& 0.21199 \\
RR Leo	& -1.60	& 0.452403	& 0.37005	& 0.59376	& 0.89671	& 0.15814	& 0.49317	& 0.82571 \\
SV Hya	& -1.22	& 0.478527	& 0.45564	& 0.90954	& 0.34279	& 0.28423	& 0.06210	& 0.70342 \\
XX And	& -1.94	& 0.722747	& 0.00464	& 0.16971	& 0.18725	& 0.68272	& 0.03730	& 0.35984 \\
SW And	& -0.07	& 0.442279	& 0.11762	& 0.38491	& 0.70367	& 0.41334	& 0.20898	& 0.15671 \\
\hline
UU Cet	& -1.00	& 0.606081	& 0.42479	& 0.78891	& 0.21132	& 0.49441	& 0.33841	& 0.79084 \\
W Tuc   & -1.64	& 0.642247	& 0.52457	& 0.08133	& 0.55307	& 0.05375	& 0.54489	& 0.84282 \\	
    \hline
    \end{tabular}
    \caption{{\bf Period (in days), [Fe/H] and $\bf \phi_1$ to $\bf \phi_6$} of the Fourier decompositions
    for {\em TESS} data for a selection of RR Lyrae stars. UU Cet and W Tuc were not included in the model building.}
    \label{tab:fourier}
\end{table}

\allauthors
\end{document}